\documentclass[pre,twocolumn,showpacs,preprintnumbers,amsmath,amssymb,byrevtex]{revtex4}
\usepackage{graphicx}% Include figure files
\usepackage{dcolumn}% Align table columns on decimal point
\usepackage{bm}% bold math

\begin{document}

%\preprint{APS/123-QED}
\title{Rotational Brownian motion on the sphere surface and rotational relaxation}

\author{Ekrem Ayd\i ner} \email{ekrem.aydiner@deu.edu.tr}
\affiliation{Dokuz Eyl\"{u}l University, Department of Physics
35160 \.{I}zmir, Turkey}
\date{\today}

\begin{abstract}
The spatial components of the autocorrelation function of
noninteracting dipoles are analytically obtained in terms of
rotational Brownian motion on the surface of a unit sphere using
multi-level jumping formalism based on Debye's rotational relaxation
model, and the rotational relaxation functions are evaluated.
\end{abstract}

\pacs{77.22.Gm, 05.40.Jc, 05.40.-a}% PACS, the Physics and Astronomy
\maketitle

Rotational Brownian motion and rotational relaxation phenomena has
received considerable interest in the literature for a long time.
Historically, the problem of isotropic rotational Brownian motion
arose in connection with the work of Purcell on spin relaxation in
systems with more than two spins in the same molecule. The dynamics
of rotational Brownian motion of a sphere around a given axis was
discussed briefly by Einstein \cite{Einstein} in one of his early
papers. Later, Debye \cite{Debye} proceeded by extending Einstein's
treatment of the translational Brownian motion to rotational
Brownian motion of noninteracting permanent dipoles in the presence
of an external time-varying field and solving the appropriate
Fokker-Planck equation. Here that equation is the Smoluchowski
equation, which is an approximate Fokker-Planck equation in the
space of angular coordinates for the distribution function of the
orientations of the dipoles on the surface of the unit sphere when
the influence of the inertia of the molecules on the relaxation
process is ignored. Following years many generalizations of Debye
model have been extensively studied
\cite{Furry,Pines,Fatuzzo,Ford,Favro,Pauling,Dattagupta1,McConnell,Brown}.
However, Debye theory cannot explain relaxation phenomena in
strongly interacting systems. The relaxation behavior in strongly
interacting systems such as amorphous polymers, spin glasses, glass
forming liquids, etc., deviates considerably from the exponential
(Debye) pattern $\Phi  \sim \exp\left[-t/\tau\right]$ and is
characterized by a broad distribution of relaxation times. The
relaxation in strongly interacting systems obey to the stretched
exponential relaxation function experimentally $\Phi  \sim
\exp\left[-(t/\tau)^{\alpha}\right]$ with $0<\alpha<1$, which often
referred to as the Kohlrausch-William-Watts (KWW) function
\cite{Kohlrausch,Williams}.

In this Letter our aim is to obtain rotational relaxation function
in terms of autocorrelation function of rotational Brownian motion
using multi-level jumping process (MJP) formalism based on
historical Debye rotational relaxation scenario. The MJP is the
generalization of the two-level jumping process (TJP) formalism. The
TJP and MJP have been used to obtain the relaxation function
\cite{Dattagupta2,Dattagupta3,Kubo,Anderson}. Recently, the
non-exponential relaxation function in complex systems has been
obtained using MJP formalism \cite{Aydiner}.

Autocorrelation function $\phi$ is simply given by
\begin{equation}\label{1}
\phi\left(t\right) =\left\langle
\mu\left(0\right) \mu\left( t\right)  \right\rangle
\end{equation}
where $\mu(0)$ and $\mu(t)$ denote values of fluctuation variable
$\mu$, i.e. dipole, at time zero and the same variable at a later
time $t$, respectively. Physically, it measures the time over which
the variable $\mu$ retains its own memory until this memory is
averaged out by statistical randomness.

If it is assumed that the external field has been applied in the
$z$-direction, the spatial components of the autocorrelation
function of dipoles can be written in terms of the Brownian motion
on the surface of the unit sphere in a simple form
\begin{subequations}\label{2}
\begin{equation}
\phi_{x}\left(  t\right)  =\left\langle \mu_{x}\left(  0\right)  \mu
_{x}\left(  t\right)  \right\rangle =\left\langle
\mu_{0}\sin\theta\cos \varphi\right\rangle
\end{equation}
\begin{equation}
\phi_{z}\left(  t\right)  =\left\langle \mu_{y}\left(  0\right)  \mu
_{y}\left(  t\right)  \right\rangle =\left\langle
\mu_{0}\sin\theta\sin \varphi\right\rangle
\end{equation}
\begin{equation}
\phi_{z}\left(  t\right)  =\left\langle \mu_{z}\left(  0\right)  \mu
_{z}\left(  t\right)  \right\rangle =\left\langle \mu_{0}\cos\theta
\right\rangle
\end{equation}
\end{subequations}
where $\mu_{0}$ denotes magnitude of the dipole ($\mu_{0}=1$),
$\theta$ and $\varphi$ are stochastic variables which define the
spatial orientation of a dipole, in other words, position of the
Brownian particle on the sphere surface. For simplicity, it can be
considered that both of $\theta$ and $\varphi$ are discrete
stochastic variables. The $\theta$ takes random values in between
$0$ and $\pi$, and on the other hand, $\varphi$ takes random values
in between $0$ and $2\pi$, where $\pi$'s value is defined in terms
of degree. Hence, stochastic variable $\mu$ can take randomly
integer or non-integer values depending on $\theta$ and $\varphi$.
In a such process, fluctuation variable $\mu$ jumps from one value
of $\mu$ to another stochastically. As it can be seen in Eqs.\,(2)
only $z$-component of the autocorrelation function depends on
$\theta$, however, $x$ and $y$ components depend on two variables
i.e. $\theta$ and $\varphi$. We will see below that all components
of the autocorrelation function are carried out using MJP after a
little bit of algebra.

All components of the autocorrelation function in Eqs.\,(2) can be
derived by using stochastic formalism. For example, the
$z$-component $\phi_{z}\left(t\right)$ of the autocorrelation
function is given by
\begin{equation} \label{3}
\phi_{z}\left(  t\right) =\sum_{\mu_{z0}}\sum_{\mu_{z}}p\left(
\mu_{z0}\right) \mu_{z0}P\left( \mu_{z0}\mid \mu_{z},t\right)
\mu_{z}
\end{equation}
where the quantities $\mu_{z0}$ and $\mu_{z}$ are the values of
stochastic variable $\mu_{z}$ at times $0$ and $t$, respectively.
Having started at the initial state $\mu_{z0}$ with the statistical
weight $p\left(\mu_{z0}\right)$, and $P\left(\mu_{z0}\mid
\mu_{z},t\right)$ measures the probability of propagation from
$\mu_{z0}$ to $\mu_{z}$ in time $t$, which is known as conditional
probability.

The conditional probability $P\left(\mu_{z0}\mid \mu_{z},t\right)$
can be given by the matrix representation associate a stochastic
state $\left\vert \mu_{z}\right\rangle$. In the operator formalism
(See Refs.\,\cite{Dattagupta2,vanKampen}) the conditional
probability $P\left( \mu_{z0}\mid \mu_{z},t\right)$ is simply
written with the matrix elements of operators as
\begin{equation} \label{4}
P\left( \mu_{z0}\mid \mu_{z},t\right)  \equiv\left\langle \mu_{z}
\right\vert \widehat {P}\left(  t\right)  \left\vert
\mu_{z0}\right\rangle
\end{equation}
in the matrix form where $\widehat{P}(t)$ is the conditional
probability operator, and $\left\vert ...\right\rangle$ indicates
vector space of stochastic states. Note that the conditional
probability satisfy
\begin{equation} \label{5}
P\left(\mu_{z0}\mid \mu_{z},0\right)  \equiv\left\langle \mu_{z}
\right\vert \widehat {P}\left(0\right)  \left\vert
\mu_{z0}\right\rangle =\delta\left( \mu_{z}-\mu_{z0}\right)
\end{equation}
for $t=0$.

Similarly, the fluctuation variable $\mu_{z}$ can be also given by
matrix form, the eigenvalues of which correspond to stochastic
variables $\mu_{0}$, $\mu_{1}$,...,$\mu_{\pi}$, respectively. The
matrix representation the non-interacting fluctuation variable
$\mu_{z}$ can be represented associated with operator $\widehat{Z}$
as
\begin{equation} \label{6}
\widehat{Z}=\left(
\begin{array}
[c]{ccccccc}%
\mu_{0} & 0 & 0 &. & . & . & 0\\
0 & \mu_{1} & 0 &. & . & . & 0\\
0 &  0 & \mu_{2} &. & . & . & 0\\
. & . & . & . & . & . & .\\
. & . & . & . & . & . & .\\
. & . & . &. & . & . & \text{.}\\
0 & 0 & 0 &. & . & . & \mu_{\pi}%
\end{array}
\right).
\end{equation}
Eigenvalues of operator $\widehat{Z}$ are given by
$\mu_{z}=$cos$\theta$ where $z=0,1,2,...,\pi$.

Thus, the autocorrelation function in Eq.\,(3) can be transformed to
the matrix representation
\begin{equation} \label{7}
\phi_{z}\left(t\right)
=\sum_{\mu_{z},\mu_{z0}}p(\mu_{z0})\left\langle \mu_{z0}\right\vert
\widehat{Z}\left\vert \mu_{z0}\right\rangle \left\langle \mu_{z}
\right\vert \widehat{P}\left( t\right) \left\vert
z_{z0}\right\rangle \left\langle \mu_{z} \right\vert
\widehat{Z}\left\vert \mu_{z} \right\rangle
\end{equation}
using above definitions.

As yet, the autocorrelation function (3) has been given by matrix
representation as seen in Eq.\,(7). The Eq.\,(7) can be solved
easily using matrix algebra and with more convenient basis vector
notation if the operator form of the conditional probability is
known.

The operator form of the conditional probability can be evaluated
from master equation. The conditional probability
$P\left(\mu_{z0}\mid \mu_{z},t\right)$ can be given associated with
master equation
\begin{equation} \label{8}
\frac{\partial}{\partial t}P\left( \mu_{z0}\mid \mu_{z},t\right)
=\sum_{\mu_{z}^{\prime }}W\left( \mu_{z}^{\prime }\mid
\mu_{z}\right) P\left( \mu_{z0}\mid \mu_{z}^{\prime },t^{\prime
}\right)
\end{equation}
where $W\left( \mu_{z}^{\prime }\mid \mu_{z}\right)$ is the jump
rate from $\mu_{z}$ to $\mu_{z}^{\prime}$ of the Brownian particle
i.e. dipole.

The jumping rate $W\left( \mu_{z}^{\prime }\mid \mu_{z}\right)$ can
be given by matrix form
\begin{equation} \label{9}
W\left( \mu_{z}^{\prime}\mid \mu_{z}\right)=\left\langle \mu_{z}
\right\vert \widehat{W}\left\vert \mu_{z}^{\prime}\right\rangle
\end{equation}
where $\widehat{W}$ is the jumping rate operator. Thus Eq.\,(8) can
be rewritten completely by a matrix form as
\begin{equation} \label{10}
\frac{\partial} {\partial t} \left\langle \mu_{z} \right\vert
\widehat{P}\left(t\right)\left\vert \mu_{z0}\right\rangle
=\sum_{\mu_{z}^{\prime}} \left\langle \mu_{z} \right\vert
\widehat{W}\left \vert \mu_{z}^{\prime}\right\rangle \left\langle
\mu_{z}^{\prime}\right\vert \widehat{P}\left(t\right)\left\vert \mu_{z0}%
\right\rangle
\end{equation}
using Eq.\,(4) and (9).

The stochastic states taken to form an orthonormal set which provide
\textit{closure} property is as
\begin{equation} \label{11}
\sum_{\mu_{z}}\left\vert \mu_{z}\right\rangle \left\langle \mu_{z}
\right\vert =1
\end{equation}
for discrete variables. Using closure property,  Eq.\,(10) is
reduced to
\begin{equation} \label{12}
\left\langle \mu_{z} \right\vert \frac{\partial
\widehat{P}\left(t\right) }{\partial t} \left\vert
\mu_{z0}\right\rangle = \left\langle
\mu_{z} \right\vert \widehat{W} \widehat{P}\left(t\right) \left\vert \mu_{z0}%
\right\rangle \ \ .
\end{equation}
This equation can be simplified as
\begin{equation} \label{13}
\frac{\partial}{\partial t}\widehat{P}\left(t\right)= \widehat{W}
\widehat{P}\left(t\right) \ \ .
\end{equation}
As a result, the solution of Eq.\,(13) is
\begin{equation}\label{14}
\widehat{P}\left(  t\right) =\exp\left( \widehat{W}t\right)
\end{equation}
where $\widehat{W}$ is the jump matrix and can be written in terms
of collision and unit matrix elements as
\begin{equation} \label{15}
\widehat{W}=\lambda\left( \widehat{J}-\textbf{1}\right)
\end{equation}
where $\lambda$ is defined as the relaxation rate, $\widehat{J}$ is
the collision matrix and \textbf{1} is a unit matrix. The relaxation
rate  $\lambda$ for $\pi$-level jumping is given by
\begin{equation} \label{16}
\lambda = \pi w
\end{equation}
and collision matrix $\widehat{J}$ is presented by
\begin{equation} \label{17}
\widehat{J}=\left(
\begin{array}{cccccc}
1/\pi & 1/\pi & . & . & . & 1/\pi \\
1/\pi & 1/\pi & . & . & . & 1/\pi \\
. & . & . & . & . & . \\
. & . & . & . & . & . \\
. & . & . & . & . & . \\
1/\pi & 1/\pi & . & . & . & 1/\pi%
\end{array}%
\right)
\end{equation}
where $w$ is the jump rate from one value of $\mu_{z}$ to another.

It is helpful to write down the explicit form of the jump matrix
$\widehat{W}$ in terms of $w$. If Eqs.\,(16) and (17) are inserted
into Eq.\,(15), the jump matrix $\widehat{W}$ can be constructed as
\begin{equation} \label{18}
\widehat{W}=\left(
\begin{array}
[c]{cccccc}%
-(\pi-1) w & w & . & . & . & w\\
w & -(\pi-1) w & . & . & . & w\\
. & . & . & . & . & .\\
. & . & . & . & . & .\\
. & . & . & . & . & .\\
w & w & . & . & . & -(\pi-1)w
\end{array}\right) \ \ .
\end{equation}

The usefulness of the decomposition in Eq.\,(15) comes out from the
fact that the $\widehat{J}$ matrix has a very simple property
\begin{equation} \label{19}
\widehat{J^{2}}=\widehat{J}, \ \ \widehat{J^{3}}=\widehat{J}, \ \
...,\ \ \ \widehat{J^{k}}=\widehat{J}
\end{equation}
for any integer $k>0$. This property allows us to immediately
construct the conditional probability matrix as
\begin{equation} \label{20}
\widehat{P}\left( t\right) =\exp \left[ \lambda \left(
\widehat{J}-\textbf{1}\right) t\right]
\end{equation}
or, alternatively, it can be given in the matrix form
\begin{equation} \label{21}
\widehat{P}\left( t\right) =\left(
\begin{array}{cccccc}
\eta +e^{-\lambda t} & \eta  & . & . & . & \eta  \\
\eta  & \eta +e^{-\lambda t} & . & . & . & \eta  \\
. & . & . & . & . & . \\
. & . & . & . & . & . \\
. & . & . & . & . & . \\
\eta  & \eta  & . & . & . & \eta +e^{-\lambda t}%
\end{array}%
\right)
\end{equation}
where $\eta$ is defined by
\begin{equation} \label{22}
\eta =\frac{1}{\pi}\left( 1-e^{-\lambda t}\right) \ \ .
\end{equation}
The conditional probability operator $\widehat{P}\left( t\right)$
can be expressed in useful form as
\begin{equation}\label{23}
\widehat{P}\left(  t\right)  =\exp\left(  -\lambda t\right)
\left[ 1-\widehat{J}+\widehat{J}\exp\left(  \lambda t\right)
\right]
\end{equation}
using direct power series expansion and relation which is given by
Eq.\,(19).

On the other hand, it is convenient to introduce the more general
notation for basis vectors to solve Eq.\,(7). Therefore, let us
associate a stochastic state $\left\vert \theta \right\rangle$
($\theta=0,1,2,...,\pi$ for multi-level jumping process) instead
$\left\vert \mu_{z} \right\rangle$ of $\pi$ values
\begin{equation} \label{24}
\left\vert 0\right\rangle =\left(
\begin{array}
[c]{c}%
1\\
0\\
0\\
.\\
.\\
.\\
0
\end{array}
\right)  ,\left\vert 1\right\rangle =\left(
\begin{array}
[c]{c}%
0\\
1\\
0\\
.\\
.\\
.\\
0
\end{array}
\right),...,\left\vert \pi\right\rangle =\left(
\begin{array}
[c]{c}%
0\\
0\\
0\\
.\\
.\\
.\\
1
\end{array}
\right)
\end{equation}

The \textit{closure} property  for new stochastic states $\left\vert
\theta \right\rangle$ (and similarly $\left\vert
\theta^{\prime}\right\rangle$) is presented as
\begin{equation} \label{25}
\sum_{\theta}\left\vert \theta \right\rangle \left\langle \theta
\right\vert =1
\end{equation}
in a discrete form.

Now, Eq.\,(7) can be rewritten using new basis vectors as
\begin{equation} \label{26}
\phi_{z}\left(  t\right)
=\sum_{\theta,\theta^{\prime}}p_{\theta}\left\langle
\theta\right\vert \widehat{Z}\left\vert \theta\right\rangle
\left\langle \theta^{\prime}\right\vert \widehat{P}\left( t\right)
\left\vert \theta\right\rangle \left\langle
\theta^{\prime}\right\vert \widehat{Z}\left\vert
\theta^{\prime}\right\rangle
\end{equation}
with $a priori$ occupation probability
\begin{equation} \label{27}
p_{\theta}=\frac{1}{\pi}, \ \ \ \ \theta=0,1,2,...,\pi \ \ .
\end{equation}

The expectation value of conditional probability matrix
$\widehat{P}\left(  t\right)$ in Eq.\,(26) is obtained as
\begin{equation} \label{28}
\left\langle \theta\right\vert \widehat{P}\left(  t\right)
\left\vert \theta^{\prime}\right\rangle =\frac{1}{\pi}+\left(
\delta_{\theta \theta^{\prime}}-\frac{1}{\pi}\right) \exp\left(
-\lambda t \right)
\end{equation}
using Eq.\,(23) and new basis vectors in Eq.\,(24), and on the other
hand, the expectation value of $\widehat{Z}$ is given by
\begin{equation} \label{29}
\left\langle \theta \right\vert \widehat{Z}\left\vert
\theta^{\prime}\right\rangle =Z_{\theta}\delta_{\theta
\theta^{\prime}}
\end{equation}
with $\theta,\theta^{\prime}=0,1,2,...,\pi$, where $Z_{\theta}$ are
the allowed values of the stochastic variables.

If Eqs.\,(27)-(29) are inserted into Eq.\,(26), the autocorrelation
function (26) can be reduced to a simple form
\begin{equation} \label{30}
\phi_{z}\left( t\right) =\left\langle \mu_{z} \right\rangle
^{2}+\left( \left\langle \mu_{z}^{2}\right\rangle -\left\langle
\mu_{z} \right\rangle ^{2}\right) \exp\left(  -\lambda t \right)
\end{equation}
where the deterministic quantities are weighted averages over the
available states of the corresponding variable. The average values
of $\left\langle \mu_{z}\right\rangle$ and $\left\langle
\mu_{z}^{2}\right\rangle$ in the stationary state are given by
\begin{equation} \label{31}
\left\langle \mu_{z}\right\rangle
=\sum_{\theta=0}^{\pi}p_{\theta}\left\langle \theta\right\vert
\widehat{Z}\left\vert \theta\right\rangle
=\frac{1}{\pi}\sum_{\theta=0}^{\pi}Z_{\theta}
\end{equation}
and
\begin{equation} \label{32}
\left\langle \mu_{z}^{2}\right\rangle
=\sum_{\theta=0}^{\pi}p_{n}\left\langle \theta\right\vert
\widehat{Z^{2}}\left\vert \theta\right\rangle =\frac{1}{\pi}\sum
_{\theta=0}^{\pi}Z_{\theta}^{2}
\end{equation}
respectively. The diagonal elements of $\widehat{Z}$ matrix (6) are
given as $\mu_{z}=\cos\theta$, i.e., $Z_{\theta}=\cos\theta$. Thus,
it is expected that $\left\langle \mu_{z} \right\rangle=0$ and
$\left\langle \mu_{z}^2\right\rangle =1/2$.

Finally, $z$-component of the autocorrelation function for the
Brownian motion on the spherical surface can be obtained simply as
\begin{equation} \label{33}
\phi_{z}\left(  t\right) = \frac{1}{2} \exp\left[ -\lambda t \right]
\ \ .
\end{equation}
As it can be seen in Eq.\,(33) $z$-component of the rotational
autocorrelation function has exponential form which indicates that
the same component of rotational relaxation must has exponential
form.

The relaxation function for $z$-component can be immediately
obtained using linear response theory
\begin{equation} \label{34}
\Phi\left(t\right) =\Psi\left(  t=\infty\right) -\Psi\left(
t\right)
\end{equation}
where $\Psi\left(t\right)$ is well known response function which can
be associated to the famous response-relaxation function as
\begin{equation} \label{35}
\Psi\left(  t\right)  =\frac{1}{kT} \left( \left\langle
\mu_{z}^{2}\right\rangle -\left\langle \mu_{z} \left( 0\right)
\mu_{z}\left( t\right) \right\rangle \right)
\end{equation}
where $k$ is the Boltzmann constant, and $T$ is the temperature. If
Eq.\,(35) is inserted into Eq.\,(34), as a final result, the
rotational relaxation function can be obtained as
\begin{equation}\label{36}
\Phi_{z}\left(t\right)=\frac{1}{2kT} \exp\left[  -\lambda t \right]
\ \ .
\end{equation}

We have shown that the $z$-component of the autocorrelation function
and relaxation function can be obtained using the multi-level
jumping formalism. Similarly, the other components of the
autocorrelation function, that is, $\phi_{x}\left(t\right)$ and
$\phi_{y}\left(t\right)$ can be also derived using this technique.
However, to obtain the components $\phi_{x}$ and $\phi_{y}$, we
should eliminate the $\varphi$ in $\phi_{x}$ and $\phi_{y}$.

At this point, firstly, it can be suggested that components of the
autocorrelation function in Eq.\,(2) must satisfy the relation
\begin{equation}\label{37}
\left\langle \phi_{x}^{2}(t)\right\rangle +\left\langle
\phi_{y}^{2}(t)\right\rangle+\left\langle
\phi_{z}^{2}(t)\right\rangle=1
\end{equation}
which is caused from a simple trigonometric identity as
\begin{equation}\label{38}
\left\langle \sin^{2}\theta\cos^{2}\varphi\right\rangle
+\left\langle \sin^{2}\theta
\sin^{2}\varphi\right\rangle+\left\langle
\cos^{2}\theta\right\rangle=1 \ \ .
\end{equation}
Therefore, the components $\phi_{x}$ and $\phi_{y}$ can be written
as
\begin{equation}\label{39}
\left\langle \phi_{x}^{2}\left(t\right)  +\phi_{y}^{2}\left(
t\right)\right\rangle = 1- \left\langle \cos^{2}\theta\right\rangle
\ \ .
\end{equation}
On the other hand, it is expected that the component
$\phi_{y}\left(t\right)$ should be equal to
$\phi_{x}\left(t\right)$. Thus, $\phi_{x}\left(t\right)$ is written
as
\begin{equation}\label{40}
\phi_{x}\left( t\right)\equiv\left\langle \phi_{x}\left(t\right)\right\rangle =\frac{1}{\sqrt{2}%
}\left\langle \sin\theta\right\rangle
\end{equation}
in simple form.

As it can be seen in Eq.\,(40), using mathematical trick, we have
got rid of $\varphi$ in the components $\phi_{x}\left(t\right)$ and
$\phi_{y}\left(t\right)$. Now, to calculate the component
$\phi_{x}\left(t\right)$ we need only values $\left\langle
\mu_{x}\right\rangle$ and $\left\langle \mu_{x}^{2}\right\rangle$.
The averages can be given as
\begin{equation} \label{41}
\left\langle \mu_{x}\right\rangle
=\sum_{\theta=0}^{\pi}p_{\theta}\left\langle \theta\right\vert
\widehat{X}\left\vert \theta\right\rangle
=\frac{1}{\pi}\sum_{\theta=0}^{\pi}X_{\theta}
\end{equation}
and
\begin{equation} \label{42}
\left\langle \mu_{x}^{2}\right\rangle
=\sum_{\theta=0}^{\pi}p_{\theta}\left\langle \theta\right\vert
\widehat{X^{2}}\left\vert \theta\right\rangle =\frac{1}{\pi}\sum
_{\theta=0}^{\pi}X_{\theta}^{2}
\end{equation}
respectively. The matrix $\widehat{X}$ has the same form with (6).
But elements of $\widehat{X}$ matrix are equal to
$\mu_{x}=\sin\theta$, i.e., $X_{\theta}=\sin\theta$. Therefore,
required averages can be carried out as $\left\langle
\mu_{x}\right\rangle =\sqrt{2}/\pi$ and $\left\langle
\mu_{x}^{2}\right\rangle =1/4$. Hence, the relaxation function for
$x$ and $y$ components are given by
\begin{equation}\label{43}
\Phi_{x}\left(t\right)=\Phi_{y}\left(t\right)=\frac{1}{kT}\left(
\frac{1}{4}-\frac{2}{\pi^2}\right)  \exp\left[ -\lambda t\right] \ \
.
\end{equation}
The Eq.\,(43) clearly show that the components
$\phi_{x}\left(t\right)$ and $\phi_{y}\left(t\right)$ of rotational
relaxation function have also exponential form.

In this Letter, the spatial components of the autocorrelation
function of noninteracting dipoles have been analytically obtained
in terms of rotational Brownian motion on the surface of a unit
sphere using multi-level jumping formalism based on Debye's
rotational relaxation model, and the rotational relaxation functions
have been evaluated. As it is expected, all components of the
rotational relaxation function have exponential form when the
influence of the inertia of the dipoles is ignored.

This work was partially supported by Dokuz Eyl\"{u}l University
(Project No: 04.KB.FEN.098)

%\begin{acknowledgments}
%\end{acknowledgments}
%\newpage %Just because of unusual number of tables stacked at end

\end{document}